\documentclass[usenatbib]{mn2e}

\newcommand\nice[1]{#1}    \newcommand\subm[1]{}   

\nice{\newcommand\dosingle[1]{#1}  \newcommand\dodouble[1]{ }} 
\subm{\newcommand\dosingle[1]{ }   \newcommand\dodouble[1]{#1}} 

\dodouble{\documentclass[usenatbib,referee]{mn2e}}
\newcommand\postrefereechanges[1]{{ #1}} \newcommand\postrefereestart{ }  \newcommand\postrefereestop{ }  

\usepackage{graphicx}  

\usepackage{amsfonts}

\usepackage{mn_bmath} 

 \usepackage{hyperref}
\nice{ 
\providecommand{\eprint}[1]{\href{http://arxiv.org/abs/#1}{{\tt [arXiv:#1]}}
}

\providecommand{\url}[1]{\href{#1}{#1}}
}

\providecommand{\adsurl}[1]{} 

\def\SSS{Sect.~}

\providecommand\apj{ApJ}                 
\providecommand\aap{A\&A}            
\providecommand\mnras{MNRAS}
\providecommand\cqg{CQG}
\providecommand\grg{Gen. Rev. Grav.} 
\providecommand\pra{Phys.~Rev.~A}
\providecommand\prd{Phys.~Rev.~D}
\providecommand\physrep{Phys. Rep.}
\providecommand\BASI{Bull. Astr. Soc. India}
\providecommand\AmJPhys{Am. J. Phys.}

\nice{ \newcommand\mycaptionfont{\protect\footnotesize} }

\def\gtapprox{\,\lower.6ex\hbox{$\buildrel >\over \sim$} \, }
\def\ltapprox{\,\lower.6ex\hbox{$\buildrel <\over \sim$} \, }
\def\propapprox{\,\lower.6ex\hbox{$\buildrel \propto\over \sim$} \, }

\def\arcs{\ifmmode {'' }\else $'' $\fi}     
\def\arcm{\ifmmode {' }\else $' $\fi}       

\def\fr7{7$ \hskip -0.9ex \vrule height0.8ex width0.8ex depth-0.73ex
                                                     \hskip0.1ex$}
\def\frtoday{Le\space\number\day\space\ifcase\month\or
  janvier\or f\'evrier\or mars\or avril\or mai\or juin\or
  juillet\or ao\^ut\or septembre\or octobre\or novembre\or 
d\'ecembre\fi\space \number\year}

\newcommand\hMpc{\mbox{$h^{-1}$ Mpc}}
\newcommand\hGpc{\mbox{$h^{-1}$ Gpc}}

\newcommand\rinj{r_{\mathrm{inj}}}  
\newcommand\Omm{\Omega_{\mathrm{m}}}
\newcommand\Omtot{\Omega_{\mathrm{tot}}}

\newcommand\rC{R_{\mathrm{C}}}

\newtheorem{definition}{Definition}{\bf}{\rm}

\begin{document}

\title[Stationary movement]{There was movement that was stationary,
  for the four-velocity had passed around\thanks{With apologies to Andrew Barton ``Banjo'' Paterson.}}

\author[B. F. Roukema]{Boudewijn F. Roukema
\\
Toru\'n Centre for Astronomy, Nicolaus Copernicus University,
ul. Gagarina 11, 87-100 Toru\'n, Poland  
\\
}


\date{\frtoday}


\maketitle 

\begin{abstract}
{Is the Doppler interpretation of galaxy redshifts in a
  Friedmann-Lema\^{\i}tre-Robertson-Walker (FLRW) model valid in the
  context of the approach to comoving spatial sections pioneered by
  de~Sitter, Friedmann, Lema\^{\i}tre and Robertson, i.e. according to
  which the 3-manifold of comoving space is characterised by both
  its curvature and topology?}
{Holonomy transformations for flat, spherical and hyperbolic FLRW
  spatial sections are \postrefereechanges{proposed}.}
{\postrefereechanges{By quotienting a simply-connected FLRW spatial
    section by an appropriate group of holonomy transformations, the} Doppler
  interpretation in a non-expanding Minkowski space-time, obtained via
  four-velocity parallel transport along a photon path, is found to
  imply that an inertial observer is receding from herself at a speed
  greater than zero, implying contradictory world-lines.  The
  contradiction \postrefereechanges{in the multiply-connected case}
  occurs for arbitrary redshifts in the flat and spherical cases, and
  for certain large redshifts in the hyperbolic case.}
{The link between the Doppler interpretation of redshifts and cosmic
  topology can be understood physically as the link between parallel
  transport along a photon path and the fact that the comoving spatial
  geodesic corresponding to a photon's path \postrefereechanges{can
    be} a closed loop in an FLRW model of any curvature. Closed
  comoving spatial loops are fundamental to cosmic topology.}
\end{abstract}




\dodouble{\clearpage} 


\newcommand\fspacecomov{
\begin{figure}  
\centering
\includegraphics[width=8cm]{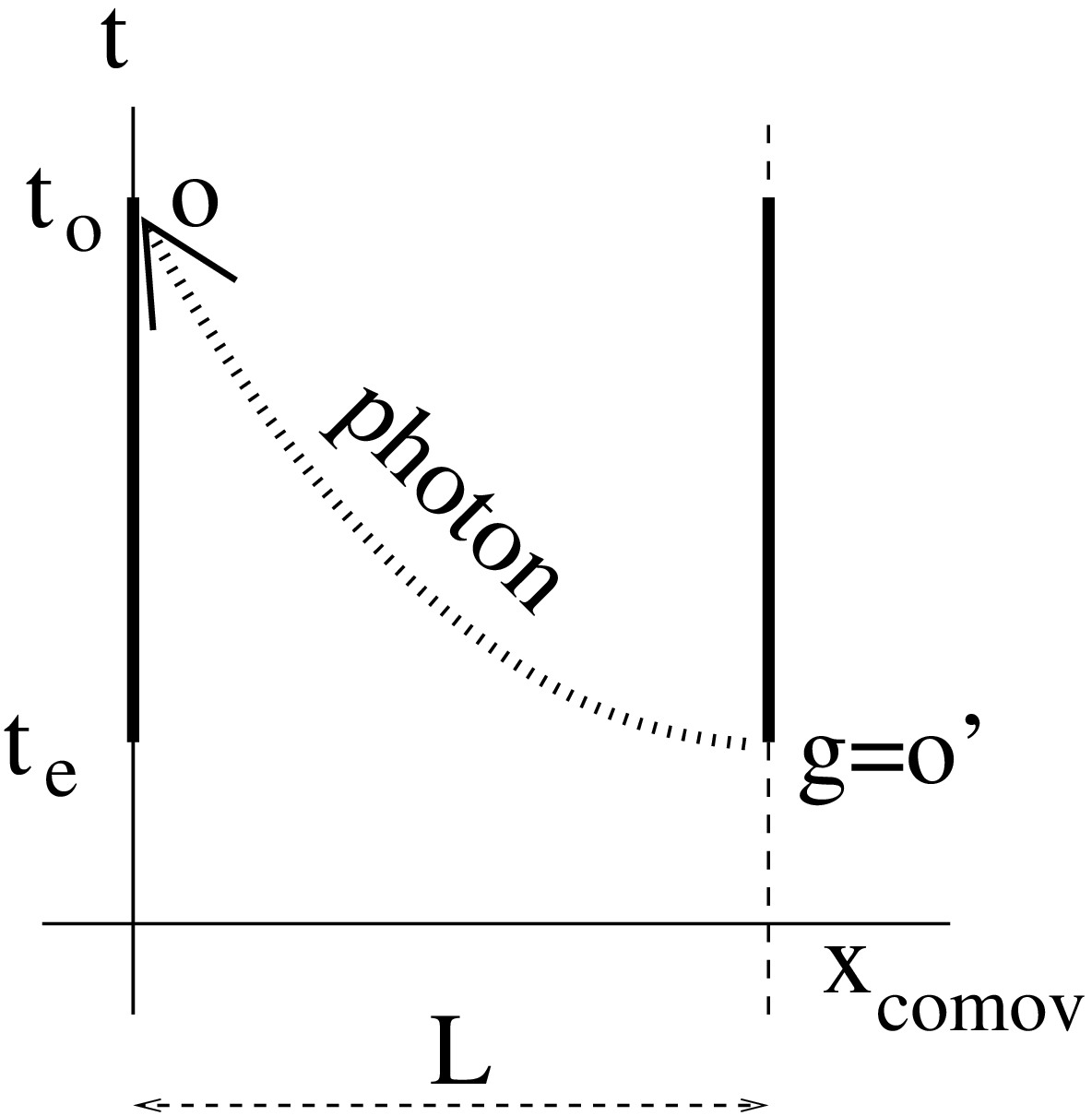}
\caption[comoving space]{ 
\mycaptionfont 
An observer ``o'' and a galaxy ``g''
both at rest in comoving coordinates 
in an FLRW model, showing one spatial dimension of the comoving covering space $\widetilde{M}$ and 
the comoving time coordinate. The galaxy emits a photon at 
cosmological time $t_{\mathrm{e}}$ that is observed by the
observer at cosmological time $t_{\mathrm{o}}$ 
to have redshift $z>0$. 
A holonomy transformation identifies the galaxy and the observer
so that the comoving space is $\widetilde{M}/\Gamma$ for a group of holonomy transformations
$\Gamma$, that varies according to curvature, as specified in 
\SSS\protect\ref{s-meth-zero}, 
\SSS\protect\ref{s-meth-pos}, and
\SSS\protect\ref{s-meth-neg}, respectively. 
The identification is exact in the zero and negatively curved cases 
(for arbitrarily selected and specially selected galaxies, respectively)
and approximate to within comoving separation $\delta$ in the positively
curved case [see Eq.~(\protect\ref{e-def-mn})].
The world-lines of the observer,
the galaxy, and the copy of the observer ``o'$\;$''
are shown as thick vertical line segments.
The world-line
of the copy of the observer coincides exactly with the world-line of
the galaxy
in the non-positively curved cases.
\label{f-spacecomov}
}
\end{figure}
} 

\newcommand\fspacedopp{
\begin{figure}  
\centering
\includegraphics[width=7cm]{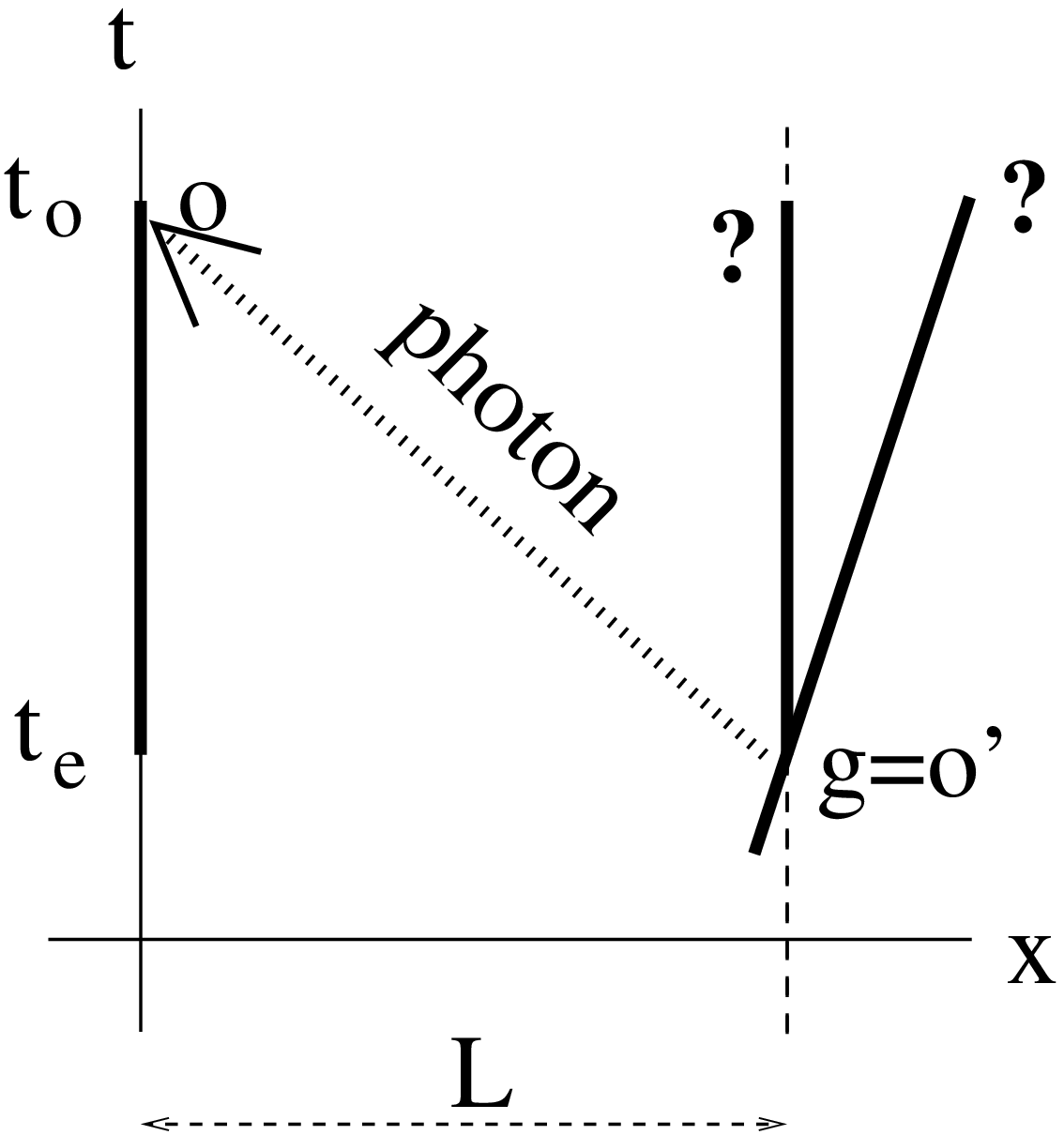}
\caption[comoving space]{ 
\mycaptionfont 
As in Fig.~\protect\ref{f-spacecomov}, shown in 
non-expanding Minkowski space-time in order to attempt to 
interpret the galaxy's redshift as a Doppler effect.
The location of the world-line of the copy of the observer 
in this diagram is ambiguous, as indicated by the question marks and discussed 
in the text (\SSS\protect\ref{s-results}): how is it possible that
the (inertial) observer is moving at constant speed away from herself 
in (non-expanding) Minkowski space-time?
\label{f-spacedopp}
}
\end{figure}
} 


\section{Introduction}
Much debate has recently taken place regarding the 
interpretation of the
redshifts of comoving galaxies in cosmological models
as a special-relativistic Doppler effect in the absence of the 
concept of expanding space
\citep{Chod07a,Barnes06,Chod07b,Abram07,Francis07,Lewis08,BunnHogg09,Peacock08,Abram09,Chod08,Faraoni09}.
%
Here, it is shown that in the flat and spherical
Friedmann-Lema\^{\i}tre-Robertson-Walker
(FLRW)
models, where galaxies are comoving massless test objects in the
standard comoving-coordinate system, the Doppler
interpretation 
in non-expanding Minkowski space-time
leads to a contradiction for galaxies at distances
from the observer that are arbitrarily small, provided that these
galaxies are considered to be comoving.  The
corresponding (weaker) argument in hyperbolic FLRW models is also presented.

In \SSS\ref{s-cos-top}, some mathematical properties
of the comoving spatial sections in FLRW models are recalled. 
The methods of using these properties in 
flat, positively curved, and negatively curved FLRW models are presented
in \SSS\ref{s-meth-zero}, \SSS\ref{s-meth-pos}, and \SSS\ref{s-meth-neg}, respectively.
The consequences are described in \SSS\ref{s-results}.
The way that density perturbations modify these consequences is
discussed in \SSS\ref{s-almostflrw}. 
The failure of the Doppler interpretation to arbitrarily small separations 
for non-negatively curved exact-FLRW models may seem to be inconsistent with the 
Minkowski nature of FLRW models in the limit towards a point in space-time.
This paradox is explained in \SSS\ref{s-Mink-limit}.
A definition of ``expansion of space'' 
and related statements 
from the literature are briefly discussed in 
\SSS\ref{s-expanding}, and the notion of an expanding Minkowski
space-time is mentioned in \SSS\ref{s-expanding-Mink}.
Conclusions are presented in \SSS\ref{s-conclu}.

For clarity, the terms ``hyperbolic'' and ``spherical'' are used for
negatively and positively curved FLRW models, respectively, rather
than the ambiguous terms ``open'' and ``closed''.  The term ``galaxy''
is used for an external massless galaxy located at some non-zero
distance from the observer in the covering space. The observer should
be considered to be located in our (massless) Galaxy. The ``galaxy''
and the ``Galaxy'' are massless in order not to violate homogeneity.
\postrefereechanges{A ``comoving spatial geodesic'' is distinct from a
  space-time geodesic. The former refers to a geodesic (a curve that
  minimises the metric distance between any close pair of points on
  that curve) of a comoving spatial section (where the metric only has
  a spatial component). It can also be thought of as a space-time
  geodesic (e.g. the path of a photon) projected to a comoving spatial
  section by ignoring the cosmological time coordinate. For example,
  in a positively curved FLRW model, a comoving spatial geodesic is an
  arc that is part of a great circle of the hypersphere $S^3$,
  since great circles are straight lines.}

\postrefereechanges{For a short introduction to the topology of FLRW
  models and observational approaches to measuring it, see
  \cite{Rouk00BASI}.  For reviews, see
  \cite{LaLu95,Lum98,Stark98,LR99,BR99,RG04}. Analyses of WMAP data
  presently include analyses suggesting that sub-matter-horizon cosmic
  topology has been detected \citep[][and references
    therein]{Gundermann2005,Caillerie07,RBG08,Aurich09a}.  and
  analyses disfavouring it \citep[][and references
    therein]{KeyCSS06,NJ07}. Key terminology regarding the comoving
  spatial section, which is a 3-manifold, includes the following,
  using the 3-torus as an example.  The manifold $M = T^3$ can be thought of
  as a cube, the ``fundamental domain'', with identified faces. It can
  also be thought of as a tiling of Euclidean 3-space, $\mathbb{R}^3$,
  called the ``covering space'', $\widetilde{M}$, which can also be
  informally
  called the ``apparent space'', since it contains many copies of any
  individual physical object. A mapping $f$ from one copy of the cube
  in $\widetilde{M}$ to another copy is a (non-arbitrary) isometry of
  $\widetilde{M}$, i.e. it preserves metric distances, and is called a
  ``holonomy transformation''. The set of holonomy transformations for
  a given manifold $M$ is the group $\Gamma$.  This group has the
  same structure as the group of all possible closed comoving spatial
  paths in $M$ that cannot be continuously transformed into one
  another. The latter group is called the first homotopy group,
  $\pi_1(M)$. Switching from $\widetilde{M}$ to $M = \widetilde{M}/\Gamma$
  can be described as ``quotienting'' $\widetilde{M}$ by $\Gamma$.}

\section{Method}  \label{s-cos-top}

The comoving spatial section of an FLRW model is a constant curvature
3-manifold whose curvature may be zero, positive or 
negative\footnote{The order $\{0,+,-\}$ is chosen for consistency 
of the presentation. This relates to the physical characteristics 
of the three types of FLRW spatial sections: 
negatively curved spaces are the
least convenient for the problem dealt with in this paper 
(see \SSS\protect\ref{s-meth-neg}).}
and whose
topology ($\pi_1$ homotopy group) may be trivial or non-trivial 
\citep{deSitt17,Fried23,Fried24,Lemaitre31ell,Rob35}. The curvature
is related to the matter-energy density through the Einstein field equations,
since the FLRW model constitutes a family of solutions to these equations. 
In contrast, so far there only exist some initial hints as to what might
eventually contribute to a theory of the topology of FLRW spatial sections
\citep[e.g., ][]{Hawking84a,Masafumi96,DowS98,CarlipSurya04,RBBSJ06,RR09}.

Nevertheless, it is clear that in an exact-FLRW model, i.e. a model
without any density perturbations of any sort, the quotienting of the
spatial comoving covering 3-manifold $\widetilde{M}$ by a group of
holonomy transformations $\Gamma$ has no effect on the metric.  The
local cosmological parameters 
(``local'' in the sense of representing a limit towards a space-time
point in an exact-FLRW model), such as the Hubble constant $H_0$ and
the total density parameter $\Omtot$, are unchanged by the
transformation from $\widetilde{M}$ to $M := \widetilde{M}/\Gamma$, so $M$ is
also an FLRW model.  

Another way of thinking about the relation between the covering space
$\widetilde{M}$ and the 3-manifold $M$ itself is that if we populate
$M$ by a set of massless galaxies, then $\widetilde{M}$ can be thought
of as the observer's {\em apparent space} containing multiple
topological images of each galaxy. Loosely speaking, $\widetilde{M}$
can be thought of as being tiled by a set of mirror copies of the fundamental domain
of $M$, except that instead of a reflection (generated by a mirror), a
holonomy transformation (an isometry that allows $\widetilde{M}$ to be
completely and exactly tiled by copies of the fundamental domain of
$M$) must be used.  The galaxies are chosen to be massless in order to avoid
violating the homogeneity condition of the FLRW models.

Consider a galaxy at redshift $z > 0$ in a simply-connected
FLRW model of comoving spatial section
$\widetilde{M} = \mathbb{R}^3$, 
$\widetilde{M} = S^3$, 
or
$\widetilde{M}= \mathbb{H}^3$,
i.e.
a galaxy that emitted photons towards the observer ``o'' when the scale factor
was $a < 1$ and the present scale factor (at the observation epoch) 
is $a_0=a_{\mathrm{o}}=1$. 
Let both the observer and the galaxy at $z$ be at rest
in comoving coordinates. Let us also assume that the observer has
been at rest in comoving coordinates since the epoch $a$.

Now let us attempt to make a Doppler interpretation of the galaxy's redshift.
A global concept of velocity is not a standard part of the FLRW model.
However, it is possible to parallel-transport the four-velocity of the galaxy
to the observer, typically along the path taken by a photon
\citep[e.g., ][]{Synge60,Narlikar94redsh,Peacock08,BunnHogg09,Faraoni09}, so that
it can be locally compared with the four-velocity of the observer. The
resulting velocity difference $\beta$ in units of the speed of light $c$ satisfies
\begin{equation}
1+z = \sqrt{\frac{1+\beta}{1-\beta}}
\label{e-doppler-z}.
\end{equation}
Does this imply that the galaxy can be thought of as receding with
velocity $\beta$ from the observer in non-expanding Minkowski space-time, and be redshifted
in the way implied by the Lorentz transformation?

\subsection{Zero spatial curvature}  \label{s-meth-zero}

In the case of zero spatial curvature of the underlying FLRW model,
let us introduce the holonomy transformation
\begin{equation}
f_0 [(x,y,z')] = (x + L,y,z'), \; \forall (x,y,z') \in \widetilde{M}
\label{e-flat-gamma}
\end{equation}
where $(x,y,z')$
is an arbitrary comoving
position\footnote{Since the position variable $z'$
is of peripheral importance, the unprimed variable $z$ is retained for the
redshift.}
in the spatial section $\widetilde{M}$, $L>0$ is the comoving
distance from the observer to the galaxy, and the orientation of
the fundamental domain is chosen so that the $x$ direction joins
the comoving positions of the observer and the galaxy.  Now 
quotient $\widetilde{M} = \mathbb{R}^3$ by 
\postrefereechanges{the group generated by $f_0$},
obtaining 
\begin{equation}
M= \mathbb{R}^3/\{ i f_0, i\in
\mathbb{Z} \} = S^1 \times \mathbb{R}^2. 
\label{e-1torus}
\end{equation}
This can be loosely referred to as a 1-torus, $T^1$ (where the 3-dimensionality
is implicit).

\postrefereechanges{The physical difference between 
$\mathbb{R}^3$ and $T^1$ models is enormous.
However, changing from the former to the latter 
does not change the metric. 
The most popular observational cosmology 
tests on various classes of
extragalactic objects cannot distinguish an $\mathbb{R}^3$ spatial
section from a $T^1$ spatial section. 
Moreover, if the Doppler interpretation of galaxy 
redshifts only depends on local, metric properties, then
it should not be affected by the change between these two 
spatial sections.}
The only change relevant to the present discussion is that 
{\em a copy of the observer is now located at the same comoving spatial position 
as the (comoving) galaxy.}
\postrefereechanges{Moreover, the ``copy'' of the observer and 
the observer herself now constitute a single physical observer.} 
Consequences of \postrefereechanges{the change from one space to
another} are considered in 
\SSS\ref{s-res-zero}.

\subsection{Positive spatial curvature}  \label{s-meth-pos}

Now consider the case where the FLRW model has
positive curvature, with radius of curvature 
\begin{equation}
\rC = \frac{c}{H_0} \frac{1}{\sqrt{\Omtot -1}}.
\label{e-defn-rc}
\end{equation}
Let us represent the comoving spatial section $\widetilde{M} = S^3$ as
a subset of Euclidean 4-space
$\{ (x,y,z',w) \in \mathbb{R}^4 : x^2 + y^2 + z'^2 + w^2 = \rC^2 \}$,
and place the observer
at $(0,0,0, \rC)$ and the galaxy at 
$[ \rC \sin(L/\rC),  0,0, \rC \cos(L/\rC) ]$, with
$L>0$ as the comoving
distance from the observer to the galaxy, as above.

Choose $m, n \in \mathbb{Z}$ such that 
\begin{equation}
0 < m < n, \; \left| \frac{2 \pi m}{n} - \frac{L}{\rC} \right| < \delta
\label{e-def-mn}
\end{equation}
for some arbitrarily small comoving distance $\delta > 0$,
since the rationals $\mathbb{Q}$ are dense in $\mathbb{R}$.
Let us introduce the holonomy transformation
\begin{eqnarray}
&&  f_+\left[\left( 
    \begin{array}{l} 
      x \\ y \\ z' \\ w 
    \end{array}
    \right) \right] = 
\nonumber \\
&&
  \left[ 
    \begin{array}{cccc} 
      \cos \theta & 0 & 0 & -\sin \theta \\ 
      0 & \cos \theta & -\sin\theta & 0 \\ 
      0 & \sin \theta &  \cos\theta & 0 \\ 
      \sin\theta & 0 & 0 &  \cos\theta
    \end{array}
    \right] 
  \;
  \left( 
  \begin{array}{l} 
    x \\ y \\ z' \\ w 
  \end{array}
  \right) 
\nonumber \\
&&
  \rule{0.5\columnwidth}{0ex} \forall (x,y,z',w) \in S^3
  \label{e-sph-gamma}
\end{eqnarray}
where $(x,y,z',w)$ are arbitrary comoving
positions in the spatial section $S^3 \subset \mathbb{R}^4$,
$\theta := 2\pi/n$,
and an arc of a great circle 
in the $x$--$w$ 2-plane joins
the comoving positions of the observer and the galaxy.  Now quotient
$S^3$ by 
\postrefereechanges{the group generated by $f_+$}, obtaining 
\begin{equation}
M= S^3/\{ i f_+, i\in \mathbb{Z} \},
\label{e-lens}
\end{equation}
which is known as the lens space L($n$,1), since the
fundamental domain is a dihedron\footnote{A polyhedron with
two (flat) faces is called a dihedron.} with two henagonal\footnote{A 
polygon with one (straight) edge is called a henagon.} (flat) faces 
\citep[e.g., ][]{GausSph01}.
This holonomy transformation, a double rotation in two orthogonal 2-planes,
has the elegant property that every
point is transported the same distance under the action of the holonomy, i.e.
it is a Clifford translation.

As in the zero curvature case, this topology change does not change 
the metric. On the other hand, what is relevant to the present
discussion is that a copy of the observer now exists at a comoving
spatial location closer than a comoving distance of $\delta > 0 $ from
the galaxy at the redshift $z$, for an arbitrarily small
$\delta$. This can be seen in the covering space by applying 
$f_+$ [the
holonomy transformation~(\ref{e-sph-gamma})]
to the observer $m$ times,
keeping in mind the choice of $m,n$~[cf. Eq.~(\ref{e-def-mn})].
Consequences of this are considered in 
\SSS\ref{s-res-pos}.

\subsection{Negative spatial curvature}  \label{s-meth-neg}

For negatively curved FLRW models, let us consider a less general
case than for non-negatively curved models. 
Consider a 
covering space $\widetilde{M} = \mathbb{H}^3$ of absolute curvature radius
$|\rC|$. 
Choose a galaxy at 
redshift $z>0$ and comoving distance $L>0$ from the observer that
satisfies
$L = 2\rinj(M)$ for an hyperbolic 3-manifold $M$ whose shortest closed
geodesic\footnote{This is a 
\protect\postrefereechanges{comoving} spatial geodesic.} is of length $2\rinj$ 
($\rinj$ is called the injectivity radius). Values of $2\rinj$ for small
hyperbolic 3-manifolds can typically be as low as $\gtapprox 0.31 |\rC|$.
For example, see table~1 of \citet{LR99} for a few small hyperbolic 3-manifolds
that could be used to find some galaxies satisfying this condition.\footnote{These
spaces could be called ``closed open'' models using the confusing
terminology frequently used in modern cosmology.}

\fspacecomov

Now orient the group of holonomy transformations $\Gamma$ that gives
$M= \mathbb{H}^3/\Gamma $ in the appropriate sense in the covering
space $\mathbb{H}^3$ so that the holonomy transformation $f_- \in
\Gamma$ associated with 
\postrefereechanges{twice the injectivity radius, i.e. $2\rinj$,} matches a
copy of the observer to the position of the galaxy, i.e. so that
\begin{equation}
f_-(\mathbf{0}) = \mathbf{x},
\end{equation}
where $\mathbf{0}$ and $\mathbf{x}$ are the observer's and galaxy's
comoving positions in $\mathbb{H}^3$, respectively.  In general, it
will not be possible to generate a 3-manifold from 
the group $\{ i f_- \;,\; i\in \mathbb{Z} \}$  
alone,
in contrast with the non-negatively curved cases. This is because of
the mathematically more complicated nature of hyperbolic 3-manifolds.
The situation is now
similar to the zero curvature case (\SSS\ref{s-meth-zero}), i.e. the
copy of the observer is located at the same comoving position as the
galaxy, with the distinction that the choice of galaxy had to be strongly
constrained to a countably infinite number\footnote{Taking into
  account the matter horizon may reduce this to a finite
  number.} of possibilities.  The consequences of changing from 
$\mathbb{H}^3$ to $\mathbb{H}^3/\Gamma$ 
are considered in
\SSS\ref{s-res-neg}.


\section{Results} \label{s-results}

\subsection{Zero spatial curvature}  \label{s-res-zero}
As explained in \SSS\ref{s-meth-zero}, changing the comoving spatial
section from $\mathbb{R}^3$ to $T^1$ does not change the metric,
and should not change the Doppler effect interpretation of
the redshift $z$ of a galaxy if that interpretation is valid.

\fspacedopp

Figure~\ref{f-spacecomov} shows the observer, the galaxy, 
the photon
emitted by the latter and absorbed by the former, 
the copy of the observer,
and the world-lines of the observer,
galaxy and copy of the observer, according to the FLRW model
of spatial section $M$, 
in the covering space $\widetilde{M}$ in comoving coordinates. 
A copy
of the observer is located at $x_{\mathrm{o'}} = L$, i.e.
at the same comoving position as the
\postrefereechanges{galaxy. Since the observer is 
comoving, the copy of the observer is also comoving, so 
the copy of the observer and
the galaxy are both comoving at the same spatial position. 
Hence, the copy of the observer and the galaxy
are} at rest with respect to one another.

It was assumed
above (\SSS\ref{s-cos-top}) that the observer has been at rest in
comoving coordinates since the epoch $a$, so the copy of
the observer has also been at rest in comoving coordinates since the
epoch $a$. Hence, the copy of the observer's world-line is 
\begin{equation}
\{(x_{\mathrm{o'}},t)\} = \{ (L,t): t_{\mathrm{e}} < t < t_{\mathrm{o}} \}.
\end{equation}
This includes the space-time event $(L,t_{\mathrm{e}})$ of the galaxy
emitting the photons that are later on observed at
$(0,t_{\mathrm{o}})$.  Hence, we can consider the copy of the observer
to have emitted photons at the same space-time event
$(L,t_{\mathrm{e}})$ in the same direction as the photons emitted by
the galaxy.  It follows that if a Doppler interpretation of the
galaxy's redshift $z>0$ is attempted based on parallel transport of
four-velocities along the photon path, leading to an inferred velocity
$\beta > 0$ [Eq.~(\ref{e-doppler-z})], then the same Doppler
interpretation implies that the copy of the observer at
$x_{\mathrm{o'}} = L$ must also be inferred to be receding at velocity
$\beta > 0$ with respect to the ``original'' observer.  However, the
observer and the copy of the observer are physically identical.  {\em Hence,
  the observer is inferred to be receding at velocity $\beta > 0$ with
  respect to herself. }

This is impossible unless the physical concept of expanding space is
added to the Doppler interpretation.  In a non-expanding Minkowski
space-time diagram for a given inertial observer, the slope of any
inertially moving object's world-line is constant and the slope
defines the object's velocity relative to that of the observer. The
slope of the observer's world-line is exactly vertical. A change of
global topology does not change this
\citep{BransStewart73,Peters83,Low90,UzanTwins02,LevinTwins01,RoukTwins06}.
An inertial observer in a non-expanding Minkowski space-time is moving
at zero velocity with respect to herself. 
She cannot recede at velocity $\beta > 0$ from herself.
Since ${x}_{\mathrm{o'}} = {x}_{\mathrm{g}} $ and
 $\dot{x}_{\mathrm{o'}} = \dot{x}_{\mathrm{g}}$, 
\postrefereechanges{there} is no
peculiar velocity difference between the galaxy and the copy of the observer
that could avoid this contradiction.

Figure~\ref{f-spacedopp} shows the equivalent information
with an attempt to support the Doppler interpretation of the galaxy's
redshift.  The galaxy is receding at a Doppler-inferred velocity of
$\beta > 0$, so in a non-expanding Minkowski space-time diagram
labelled in units with $c=1$, the galaxy
must have a world-line at an angle of $\mathrm{atan} \beta$ from the
vertical.
Since ${x}_{\mathrm{o'}} = {x}_{\mathrm{g}} $ and
$\dot{x}_{\mathrm{o'}} = \dot{x}_{\mathrm{g}} $ where $x$
is a comoving coordinate, equality must also hold in the Minkowski
diagram. That is, in Fig.~\ref{f-spacedopp},
the copy of the observer must also be receding at $\beta > 0$ from
herself, i.e. must also have a world-line at an angle of
$\mathrm{atan} \beta$ from the vertical, that coincides exactly with
the galaxy's world-line. 
However, the world-line of the galaxy and copy of the observer 
in Fig.~\ref{f-spacedopp} are labelled with a question mark,
since the
observer must be stationary with respect to herself in a non-expanding
Minkowski space-time, i.e. the copy of the observer must have a
{\em vertical} world-line.  The latter is also indicated by a question mark
in Fig.~\ref{f-spacedopp}. 
The copy of the observer cannot have two physically distinct world-lines.
The Doppler interpretation is clearly self-contradictory.

\subsection{Positive spatial curvature}  \label{s-res-pos}

The situation for spherical FLRW models (\SSS\ref{s-meth-pos}) is
nearly the same as for flat FLRW models, with the difference that the
$m$-th copy of the observer is located arbitrarily close (less than $\delta$)
from the galaxy [see Eq.(\ref{e-def-mn})],
rather than exactly at the galaxy.  The assumption that parallel transport
of a four-velocity enables inference of a Doppler recession velocity 
can be used to show that the velocity between the copy of the observer
(within $\delta$ of the galaxy) and the galaxy itself can be made arbitrarily
small. This is sufficient to show that 
the copy of the observer is receding 
at a speed $\beta^- \ltapprox \beta $ arbitrarily close to $\beta > 0$,
i.e. the observer is receding from herself at non-zero velocity $\beta^-$.
Again this is impossible in a non-expanding Minkowski space-time.

\subsection{Negative spatial curvature}  \label{s-res-neg}
The hyperbolic FLRW models (\SSS\ref{s-meth-neg}) have a much more 
restrained domain where the Doppler interpretation in non-expanding
Minkowski space-time fails. For simplicity, only galaxies 
at distances $L = 2\rinj(M)$ for an hyperbolic 3-manifold $M$
are considered in this paper. For example, with $\Omtot = 1.015, \rC = 24 {\hGpc}$,
and $\rinj \gtapprox 0.31$, this requires $z \gtapprox 21$.\footnote{For 
$\Omm=0.3, \Omega_\Lambda=0.7, h:= H_0/(100$~km/s/Mpc$)$.} 
Given this restriction, the Doppler
interpretation of the redshift of the would-be massless extremely
high redshift galaxy again
leads to a contradiction: the observer must be receding with respect
to herself, which is absurd unless an expanding space concept
is introduced.

\section{Discussion} \label{s-discuss}

\subsection{Almost-FLRW models} \label{s-almostflrw}

In the flat case, the self-contradiction of the Doppler interpretation
is limited below in $z$ only by the requirement that $z$
be large enough that the (massless) galaxy be at rest in comoving
coordinates. In an exact-FLRW model, i.e. in the absence of density
perturbations, there is no gravitational collapse of the cosmic web of 
filaments of large-scale structure, galaxies and clusters of 
galaxies, and no turnaround radius, so 
comoving ``galaxies'' exist arbitrarily close to the observer, 
at any distance $L > 0$.
In this case, even with a ``galaxy'' at $0 < z \ll 1$, the application of 
the holonomy transformation (\ref{e-flat-gamma})
will still lead to a contradiction:
the copy of the observer at the location of the comoving galaxy Doppler-interpreted
still recedes
at velocity $\beta > 0$ from herself.

However, in reality, density perturbations certainly do exist, so the FLRW
model is certainly wrong. The standard (formally inconsistent, but practical)
approach of using an FLRW model together with perturbations implies
a turnaround radius, so that the comoving coordinate system of the
FLRW model only makes physical sense for galaxies at comoving
distances $\gg 10$~{\hMpc}. Hence, in a perturbed FLRW model, 
the failure of the Doppler interpretation of galaxy redshifts occurs
only for $L \gg 10$~{\hMpc}, where $L$ is the comoving distance to the
galaxy, i.e. for $z \gg 0.003$.

The situation is similar for a perturbed spherical FLRW model. Again, 
there is some domain where the 
Doppler interpretation of cosmological redshifts is valid, and it should no longer be possible to
obtain $\delta > 0$ arbitrarily small. 
An evaluation of the domain of parameter space where there
would be no self-contradiction in the Doppler
interpretation of redshifts in non-expanding Minkowski space-time 
in the perturbed spherical FLRW case would require calculations
that depend on $\rC,$ the matter
density parameter $\Omm$, the dark energy parameter/cosmological constant
$\Omega_\Lambda$, and $z$. For some parameter combinations sufficiently far from
realistic estimates, even some redshifts $z \gg 0.003$ might be 
interpreted as a parallel-transported relativistic Doppler effect
without leading to an observer receding from herself in a non-expanding
Minkowski space-time.

On the other hand, in contrast with the flat and spherical cases, allowing
perturbations in an hyperbolic FLRW model should not avoid the self-contradiction
of the Doppler interpretation,
since the possible ``galaxies'' chosen in this case are at very high redshifts.

The requirement that an FLRW model be perturbed in order for the
Doppler interpretation to be valid at small redshifts ($z \ltapprox
0.003$) seems to be an implicit assumption in presentations such as
Sect.~II of \citet{BunnHogg09}. In a perfectly homogeneous FLRW model
with a massless police officer aiming a radar at a nearby massless,
comoving,
``speeding'' car, the road itself is stretching and the local
environment of the car is comoving with that car.  The police officer
can aim the radar at a building adjacent to the car and will infer
that the building is also receding at a recklessly high speed, e.g.,
150 km/h.  Moreover, the driver can open his door and touch the road
with his fingers, without any risk of injury, since both the car and
the patch of road in the car's immediate neighbourhood
are receding at 150 km/h. Let us now quotient the
space by a 
\postrefereechanges{group of
holonomy transformations} as described above, without
modifying the FLRW nature of the model, and the police officer
suddenly notices that it is he himself who, along with the patch of
road, is receding at a speed of 150 km/h \ldots from himself. He is reluctant
to give himself a speeding ticket, since he is convinced of his innocence---he 
is stationary with respect to his local surroundings. 
He decides that
there is something unphysical about interpreting the redshift as a
velocity, and concludes that space itself is 
\postrefereechanges{small and expanding.}

This analogy seems strange because roads are normally of fixed length
in ``physical'' units. For this to be possible, density perturbations
and gravitational collapse to structures that stabilise in physical
spatial coordinates must be added to the model. 
If perturbations are added, then small, local regions that are {\em not}
comoving are created.
However, in an exact-FLRW model
of zero or positive curvature, the contradiction as derived above
occurs, no matter how small the length scale, since ``galaxies''
at arbitrarily small distances are comoving in this idealised model.

\subsection{Is the failure of the Doppler interpretation consistent with 
the Minkowski limit of an FLRW space-time?}  \label{s-Mink-limit}

An FLRW model must necessarily approach Minkowski space-time in the 
limit towards any point in space-time.
Is this consistent with the
derivation above of a Doppler interpretation failure for arbitrarily 
small $z$, or equivalently, arbitrarily small $L$? Clearly yes, 
since 
\begin{equation}
  \forall L > 0, \exists \epsilon 
  \; : \; 
  0 < \epsilon \ll L.
\end{equation}
For any small $L$, the limit of space-time properties 
towards a point 
must be evaluated through even smaller 
neighbourhoods (e.g. 4-cylinders of size $|\delta x| < \epsilon, 
|\delta t| < \epsilon$)
around the observer. Hence, in an exactly homogeneous FLRW
model, there is no problem. In an almost-FLRW model, as mentioned above,
some restrictions to the results apply.

\subsection{Expanding space and finiteness} \label{s-expanding}
\citet{Francis07} clarify the notion of expanding space by defining it
as the physical situation where ``the distance between observers at
rest with respect to the fluid increases with time''. 
Using standard FLRW notation, 
an equivalent,
but more compact definition of ``expanding space'' is: 
\begin{definition}
``expanding space'' 
\postrefereechanges{means $\dot{a}(t) > 0$.}
\label{d-eos}
\end{definition}
\citet{Chod07b} introduces the terminology ``expansion of the 
cosmic substratum'' for this FLRW notion of expanding space.

The expansion of space becomes even clearer when \citet{Peacock08}'s
statement that ``undoubtedly space is expanding'' in the spherical
case and \citet{Chod07b}'s equivalent statement that ``the proper
volume of a [spherical] FL universe increases as $[a(\tau)]^3$; more
and more space thus appears'' are extended to early twentieth-century
results \citep{deSitt17,Fried23,Fried24,Lemaitre31ell,Rob35}\footnote{The author
is not aware of discussion of multiply-connected spatial sections 
by A.~G. Walker.} regarding
the role of the topology of the 3-spatial section of the Universe as a
property to be determined by observations.  The latter approach leads
to the present-day understanding of 3-manifolds, according to which a
flat or hyperbolic FLRW universe can have a finite spatial section,
i.e. there is no physical reason why a realisation of a constant
curvature 3-manifold should be simply connected. Hence, Peacock's and
Chodorowski's statements can be updated in the spirit of de~Sitter,
Friedmann, Lema\^{\i}tre and Robertson: {\em space is expanding in a
  compact FLRW model of zero, positive, or negative curvature during
  epochs \postrefereechanges{when $\dot{a} > 0$}.}

In other words, Peacock and Chodorowski agree that it is natural 
to see finite FLRW models as 
\postrefereechanges{(globally)}
expanding, and it is hard to imagine any
physical reason
why this should apply only to finite spherical FLRW models and not to
finite flat and hyperbolic FLRW models. 

The notion of expanding space in infinite models, e.g. 
the simply-connected non-positively curved models, is more difficult
to conceive of globally, since $\infty \not\in \mathbb{R}$ 
and even in typical extensions of $\mathbb{R}$ that include $\infty$, 
the expression $\infty - \infty$ is usually undefined. On the
other hand, a local concept of expanding space, where 
``local'' means any finite (greater than zero) region of the comoving
spatial section, i.e. a region attached to a framework of galaxies, 
gives a notion of expanding space. Such a region expands as $a^3$ 
\postrefereechanges{when $\dot{a}>0$}. 
The condition that every finite (greater than zero) region should satisfy
this in a non-compact exact-FLRW model provides an intuitive way of understanding
Defn~\ref{d-eos} in the case of an infinite comoving space. 
A minor caveat is that in order to avoid ambiguity,
in an almost-FLRW model, this use of ``local'' would have to be distinguished
from the even more local concept of regions smaller than the turnaround radius.

Nevertheless, it is the finite models (or, at least, models that are
compact in at least one direction) that are the most interesting not
only for extending the concept of expanding space 
as presented by Peacock and Chodorowski,
but also for providing the holonomy transformations used above
to lead to the contradiction of an inertial observer receding from herself.

\subsection{Would an expanding Minkowski space-time make physical sense?}
\label{s-expanding-Mink}

Although quotienting a Minkowski space-time by
\postrefereechanges{a group of 
spatial holonomy transformations}
at constant time for a preferred inertial observer retains the common sense
physical notion that an inertial observer is not moving with respect to 
herself 
\citep{BransStewart73,Peters83,Low90,UzanTwins02,LevinTwins01,RoukTwins06},
it is possible to add the notion of expanding space to 
a Minkowski space-time in order to make it an ``expanding space-time''
\citep{Peters86}.
In this case, an observer {\em can} have a non-zero velocity with
respect to herself. Hence, ironically in the context of the recent debate,
parallel-transport of four-velocities along photon paths {\em can} 
allow cosmological redshifts to be interpreted as a 
relativistic Doppler effect without the contradiction presented here,
{\em provided that the concept of expanding 
space is added to the Minkowski space-time used for this interpretation,}
and provided that the velocity is thought of as being tied to a path
and not as a global concept.

\postrefereestart
\subsection{Can assuming trivial topology avoid the contradiction?}
\label{s-assume-trivial}
In order to avoid the contradiction in the Doppler interpretation of
galaxy redshifts in non-expanding Minkowski space-time as presented
here, would it be sufficient to consider a limited spatial region
within an FLRW model, or to consider FLRW models whose spatial
sections are simply-connected (where every closed loop is continuously
contractible to a point)?

The former would be sufficient to avoid the contradiction, if defined
appropriately. If the galaxy of interest in a specific FLRW model is
closer to the observer than twice the injectivity radius,
i.e. $2\rinj$, then by the definition of the injectivity radius, a
closed spatial loop cannot occur.  Nevertheless, given the present
empirical estimates of the metric parameters of the Concordance Model,
even the most restrictive of the three curvatures, i.e. the hyperbolic
case, would still imply a contradiction that could be applicable to
the earliest galaxy building blocks. For example, with ${\Omm} =
0.30, {\Omega_\Lambda} = 0.72$, and the hyperbolic FLRW spatial
section v$2959(3,4)$ \citep{WeeksPhD85}\footnote{The 3-manifold name is
  that used in the file
  \url{/usr/share/snappea/ClosedCensusData/}
  \url{ClosedCensusInvariants.txt}
  of the Debian version 3.0d3-20 of Weeks' {\sc SnapPea} program, which
  is available under the GNU General Public Licence.}, which has
$2\rinj \approx 0.300$, the comoving length of the shortest closed
spatial geodesic would be about 6364{\hMpc}, so that it would
be possible for a closed loop from the observer to herself at $z\approx 8.30$ 
to occur, if she is appropriately positioned.

The latter possibility would partially avoid the contradiction, by
making the link between global spatial topology and the Doppler
interpretation explicit as a case to exclude.  The simply-connected
spherical case would also require a domain restriction, depending on
the values of ${\Omm}$ and ${\Omega_\Lambda}$ and the epoch of the
observer, since as the universe recollapses, a cosmological time can
occur when the observer sees a distant (blueshifted) copy of herself
approaching.

However, in either case, the apparent simplicity of the Doppler
interpretation in non-expanding Minkowski space-time would be
complicated by adding the global-topology related exclusion.  
\postrefereestop

\section{Conclusions} \label{s-conclu}

Cosmic topology is not just a luxury referred to by de~Sitter,
Friedmann, Lema\^{\i}tre and Robertson
\citep[and ][in the pre-relativistic era]{Schw00}\footnote{English translation: \protect\citet{Schw98}.}.
Apart from providing a competitor
to the infinite flat model for understanding WMAP observational data,
cosmic topology can also help to improve physical insight into 
fundamentals of FLRW models that initially appear to be unrelated to global
topology. 
\postrefereechanges{
If the parallel-transported four-velocity
Doppler interpretation of galaxy redshifts
depends only on the metric (including $a(t)$), then it}
leads to the physical
contradiction of an inertial observer in a non-expanding Minkowski 
space-time receding from herself. 
This contradiction can be resolved,
but only at the cost of introducing the notion of expanding space, in 
which case the motivation for a Doppler interpretation is weakened,
or by explicitly stating that the comoving distance to the galaxy
must be less than twice the injectivity radius of comovinig space.

The link between the Doppler interpretation of redshifts and cosmic
topology can be summarised physically as follows.
\begin{list}{(\roman{enumi})}{\usecounter{enumi}}
\item 
  The relativistic Doppler interpretation of a cosmological redshift
  is obtained by parallel transporting the emitting galaxy's four-velocity
  along a photon path to the observer.
\item 
  \postrefereechanges{For an FLRW model of any of the three curvatures
    (with the restrictions as detailed above), the comoving spatial
    geodesic of the photon's path can be considered to be a closed
    path by changing from a simply-connected FLRW model to a
    multiply-connected FLRW model, without any modification of the
    metric. This should not affect the Doppler interpretation if that
    interpretation depends only on the metric.}
\item 
  Closed comoving spatial paths constitute the mathematical foundation
  of cosmic \postrefereechanges{topology 
    \citep[e.g., \SSS{}3.4,][]{LaLu95}}.
\end{list}
Hence, it is unsurprising that insistence on the possibility of interpreting
galaxy redshifts as a relativistic Doppler effect leads to the properties
of cosmic topology, and in turn requires the physical concept of 
expanding space, i.e. epochs 
\postrefereechanges{when $\dot{a}>0$} (Defn~\ref{d-eos}).

\section*{Acknowledgments}

Thank you to Bartosz Lew for several useful comments.
\postrefereechanges{Thank you also to an anonymous referee
for several constructive comments.}



\subm{\clearpage}


%


\end{document}